\documentclass{sig-alternate}

\usepackage{caption}
\usepackage{subcaption}

\def\forh{\mbox{\tt for(i=0;i<N;i++)}}
\DeclareFontShape{OT1}{cmtt}{bx}{n}{<5><6><7><8><9><10><10.95><12><14.4><17.28><20.74><24.88>cmttb10}{}

\usepackage{array}
\makeatletter
\newcommand{\thickline}{%
    \noalign {\ifnum 0=`}\fi \hrule height 1pt
    \futurelet \reserved@a \@xhline
}
\newcolumntype{"}{@{\hskip\tabcolsep\vrule width 1pt\hskip\tabcolsep}}
\makeatother

\usepackage{listings}

\usepackage{url}

\lstdefinestyle{mm}{caption=function, mathescape,captionpos=b, basicstyle=\small\ttfamily}
\lstdefinestyle{rm}{mathescape,basicstyle=\small\ttfamily}

\renewcommand{\textfloatsep}{2pt}
-lmtk10

\def\forh{\mbox{\tt for(i=0;i<N;i++)}}
\def\forconstruct{omp\_for}
\def\secconst{omp\_sections}
\def\singlesection{single\_omp\_section}
\def\singleconstruct{omp\_single}
\def\parallel{omp\_parallel}
\def\biglittle{big.LITTLE}
\def\omp{OpenMP}
\def\master{master thread}

\def\big{big}
\def\little{LITTLE}

\def\parseg{parallel-segment}
\def\cparallel{\textbf{\#pragma omp parallel}}
\def\cfor{\textbf{\#pragma omp for}}

\def\wl{wl}
\def\wlb{wl_{big}}
\def\wll{wl_{LITTLE}}
\def\wli{wl(i)}
\def\wlj{wl(j)}

\def\perf{P(ct)}

\def\palu{P_{OP}(ct)}
\def\pmem{P_{MEM}(ct)}
\def\palub{P_{OP}(big)}
\def\pmemb{P_{MEM}(big)}
\def\palul{P_{OP}(LITTLE)}
\def\pmeml{P_{MEM}(LITTLE)}
\def\wlim{wl_{im}}
\def\nthreads{\texttt{N\_THREADS}}
\def\niters{\texttt{N\_ITRS}}
\def\status{\texttt{status}}
\def\statusshared{\texttt{SHARED}}
\def\statusprivate{\texttt{PRIVATE}}
\def\statusinitialprivate{\texttt{INITIAL\_PRIVATE}}
\def\doiter{\texttt{doitr}}
\def\endworklist{\texttt{end}}
\def\scaledendi{\texttt{scaledend\_i}}
\def\lock{\texttt{lock}}

\def\iter{\texttt{itr}}
\def\chunk{\texttt{chunk}}
\def\lockupdate{\texttt{lock\_update}}

\begin{document}

\title{Compiler Enhanced Scheduling for OpenMP for Heterogeneous
Multiprocessors}

\numberofauthors{2} 
\author{
\alignauthor
Jyothi Krishna V S\\
       \affaddr{IIT Madras }\\
       \email{jkrishna@cse.iitm.ac.in}
\alignauthor
Shankar Balachandran\\
       \affaddr{IIT Madras}\\
       \email{shankar@cse.iitm.ac.in}
}
\date{13 November 2015}

\maketitle
\begin{abstract}
Scheduling in Asymmetric Multicore Processors (AMP),
a special case of Heterogeneous Multiprocessors, 
is a widely studied topic. 
The scheduling techniques which are mostly runtime 
do not usually
consider parallel programming pattern used 
in parallel programming frameworks like \omp{}.
On the other hand, current compilers for these parallel
programming platforms are hardware
oblivious which prevent any compile time
optimization for 
platforms like \biglittle{} and has to
completely rely on runtime optimization.
In this paper we propose a hardware aware Compiler
Enhanced Scheduling (CES) where common compiler
transformations are coupled with compiler added scheduling
commands to take advantage of the hardware asymmetry and
improve the runtime efficiency.
We implement a compiler for \omp{} and demonstrate
its efficiency in Samsung Exynos with \biglittle{}
architecture. 
On an average, we see 18\% reduction in runtime and
14\% reduction in energy consumption
in standard NPB and FSU benchmarks with CES
across multiple frequencies and core configurations
in \biglittle.
\end{abstract}


\section{Introduction}

Heterogeneous multiprocessors use more than one type of 
processing elements (PEs) to build complex systems, 
typically hoping to conserve energy.
Asymmetric Multi Processors (AMPs) are a special case where 
the PEs have the same Instruction Set
Architecture (ISA) 
but may have different clock speeds, cache sizes or micro architecture.
ARM's \emph{big.LITTLE} and NVIDIA's Kal-El
are examples of AMP.

In AMP, PEs can be logically classified into memory
centric and computation centric.
The memory centric cores
have lower cache latency, but lower computation power when compared
to compute centric cores.
Consequently for different threads, based on the number of memory operations
and computations performed, the performance can vary
across these types of processors. 
Thus various scheduling patterns of
threads can yield different performance and power readings for
the same set of inputs.
Scheduling in AMP  is a  widely studied~\cite{tongli,mihaipricopi,pie_VanCraeynest,                                                     
Becchi,biasKoufaty,criticalaater} area. 
Different  scheduling mechanisms were proposed to exploit
the hardware heterogeneity for optimal performance and/or power. 

\biglittle{} is an AMP from ARM targeting
the mobile industry. 
The powerful cores are known as \big{} core and 
the weaker power-efficient  cores are known as \little{} cores. 
A comparison of  \big{} and \little{} cores in the system we use 
is given in Table~\ref{biglittle}.
Numerous scheduling algorithms have been suggested for \biglittle{}
to extract the best out of both set of cores. 
The most popular and widely accepted one is the Heterogeneous 
Multiprocessing (HMP)~\cite{hmp} scheduling which is integrated into
fair scheduling policy in Linux~\cite{cfs}
kernel. 

\begin{table}
\centering
\caption{\big{} and \little{} cores : Comparison}
\begin{tabular}{|p{2.2cm}|p{2.2cm}|p{2.4cm}|}
\hline
 & \textbf{\little{} core} & \textbf{\big{} core} \\ 
\hline
Core Types & Cortex-A7 & Cortex-A15 \\ 
\hline
Pipeline & simple 8-stage in-order & out-of-order, multi-issue \\ 
\hline
Frequency & 600 - 1300 MHz & 800 - 1900 MHz \\
\hline
Speed & 1.9 DMIPS~\cite{dmips} & 3.5-4.01 DMIPS  \\
\hline
Instruction Set & \multicolumn{2}{c|}{Thumb-2}\\
\hline
\end{tabular}
\label{biglittle}
\end{table}

In HMP scheduling implemented in 
big.LITTLE, the
scheduler uses
 \big{} cores  for compute intensive tasks and \little{} cores for less compute intensive
(or memory intensive) tasks.
 Based on the recent \emph{cpu utilization} by each thread, the threads are up-migrated 
(from \little{} to \big{}) or down-migrated (from \big{} to \little{}). 

In this work, we introduce a compiler enabled 
scheduling (CES) framework for multithreaded
programs, through program transformations to improve on the execution time 
and energy consumption.
Compiler analysis and transformations typically accomplish
optimizations by efficiently estimating the run-time behaviour
~\cite{Kadayifcompiler,ShihCompiler,McKinleycompiler}.
We chose \omp{}, as it is a widely used parallel programming platform. 
Many of the 
parallel programming design patterns are easily realizable in OpenMP.

The rest of paper is organized as follows. 
Section 2 will give a brief introduction to OpenMP.
Section 3 explains our proposed transformation.
Section 4 provides the implementation details and Section 5 provides
the results obtained. Section 6 lists out the related works
and Section 7 provides the conclusion.

\section{\omp{} API}
In this section we provide an introduction to OpenMP. OpenMP API is
designed for shared memory parallelism in C, C++ and FORTRAN.
We focus on C and C++ programs.
Special directives (\emph{\#pragma}) are used by the programmer
to specify OpenMP program behaviour.

A parallel region, written inside \emph{\#pragma omp parallel} (\parallel{}),
creates a team of threads, which may
run in parallel to execute the code defined.
The number of threads (\nthreads{}) in a team can be set using environment
variables or using \omp{} library calls.
For our experiments and transformations, we assume the
number of threads is equal to the number of available cores,
unless specified inside the input program.

The \master{} is a special thread
which creates the team on encountering the
\parallel{} region.  
It has  id 0 and the rest of the threads,
referred to as non-master threads, have  ids from 1 to \nthreads{} - 1.
Barriers (\emph{\#pragma omp barrier}) are used to synchronize among the threads in
a team. 
Threads waits at the barrier until rest of the threads in the team reach the
barrier. 
There is implicit barrier at the end of parallel regions
as well.

 \emph{Work-sharing} constructs define units of work, 
each of which is executed exactly once by
one of the threads in the team.
The work sharing constructs have an implicit barrier
at the  end, which can be removed using \emph{nowait} clause. 
There are three worksharing constructs available in C/C++. 
i) \forconstruct{} (\emph{\#pragma omp for}), 
implements a parallel for loop in \niters{} iterations, 
where each iterations (\niters{}) are executed once by
one of the threads. 
The iterations can be executed in parallel with other iterations.
The scheduling pattern of iterations to threads may be specified using static,
dynamic or guided clauses. 
In static scheduling each thread is given equal-sized chunks of
iterations in a round-robin fashion, until there are no iterations left.
The size of a chunk unless specified, will be taken as (\niters{}/\nthreads{}).
In dynamic scheduling, each thread is supplied with a chunk of iterations 
on demand by thread, until there are no iterations left.
Guided scheduling is very much similar to dynamic scheduling except that the chunk size
(whose intial value can be specified using th eoptional parameter)
+starts off large and decreases in later steps to handle the load imbalance better.
Guided and dynamic scheduling are used for handling load imbalance between iterations,
but has higher overhead compared to static scheduling.
In ii)\secconst{} (\emph{\#pragma omp sections}) we have multiple sections each encapsulated by
\emph{\#pragma omp section} (\singlesection{}), executed by
one of the threads in the team. 
The scheduling is arbitrary. 
iii) \singleconstruct{} contain a block that is executed by only one of the threads.


\section{CES}

In this section we explain our compiler enhanced scheduling (CES) framework.

The \omp{} programming model design works well for Symmetric Multicore Processor (SMP) environment.
However, the vast difference in computing power between \big{} and \little{}
cores can result in large difference in execution time among the
threads 
leading to hardware under-utilization.
The HMP scheduler is aimed at reducing this gap,
but it works solely based on history, leaving 
the scheduling oblivious of future workload.
The hardware-asymmetry aware CES compiler is aimed  at reducing this gap.
CES optimizes the parallel execution time by reducing the
disparity in individual thread running time.
Optimizations can also be drafted to  
reduce overall system power
consumption.

An \parallel{} region in the input \omp{} code is divided into separate 
\emph{\parseg{}}s.
A \parseg{} can be (i)one of the \omp{} worksharing constructs or 
(ii)other blocks inside \parallel regions bounded by barriers.
Each \parseg{} is analyzed and optimized separately.
The execution time of a \parseg{} will be determined by the longest running thread in 
the team.

For effective compile-time scheduling, we need to estimate the runtime behaviour of a \parseg{}
in each core. 
We have a workload model to estimate the performance of each thread
in a core.
Using the estimate we transform the code to normalize the execution time
for each thread.  

\subsection{Workload Modelling}
The Performance Estimation 
Mathematical (PEM) model would
give an estimate of the performance of a thread in \big{} and \little{} cores.
PEM use an adapted version of \cite{McKinleycompiler} for multithreaded programs.
Whilst execution, the CPU cycles could be spent in executing ALU or memory operations, 
or on branch mis-predictions.
The estimated performance $P$, of a core can be denoted as: 
\vspace{-1mm}
\begin{center}
 $   \perf{} =  \palu{} + \pmem{}  $ 
\end{center} 
\vspace{-1mm}
where $ct$ can be either \big{} or \little{}.
 The $\palu$ component will estimate the performance of each core for ALU  
operations from the number of arithmetic, floating point and bit-wise operations
performed in the code segment. 
$\pmem$ will give an estimate of the time spent in memory operations and branches based on
their count. 
The ratio of $\palub$ to $\palul$ will be very small
 since
big cores will execute the ALU operations much faster than \little{} cores for same code.
Whereas the ratio of $\pmemb$ to
$\pmeml$ will be much higher.
The $\perf$ is takena s the workload for thread i, $\wli$.

When there are multiple paths (due to branches), 
we take the largest cost from all paths
as the workload ($\wl$) of that thread. 
We collect the amount of memory, ALU and branch operations
that might be performed through simple compiler passes. 
Based on these metrics  
the performance of a thread in a particular core can be
estimated~\cite{McKinleycompiler}. 
The \big{} and \little{} cores are profiled extensively to identify the
effect of each hardware operation (memory, ALU, branching) 
during execution. 
The estimate of recursions and non-deterministic loops 
(while and do-while) are treated as \emph{unknown}
with high costs. 

Once the PEM is modelled, a thread is then scheduled into a core 
that suites 
its workload
with an intention to minimize the workload imbalance among threads inside
a \parseg{}. 
\vspace{-2.5mm}
\begin{center}
   $\displaystyle{\min_{i,j \in team}}$ $\wlim$ = $|\wli - \wlj|$
\end{center}
\vspace{-2.5mm}

Here $\wlim$ denotes the workload imbalance in the system.


In CES, an initial thread-to-core scheduling is fixed with 
 thread \emph{i} scheduled to core \emph{i}.
This can be changed if better (lower $\wlim$) scheduling pattern can be identified
at either compile time or runtime.

Now we explain compiler transformations for each type of \parseg{}.

\subsection{Scheduling \forconstruct{}}

One of the most important \parseg{}s which affects the performance 
is \forconstruct{}.
The current scheduling policies for the \forconstruct{}
are static, dynamic and guided. 
The static scheduling in
\omp{}, is designed to work well for balanced load, in a SMP.
In \biglittle{}, there is an intrinsic imbalance
in hardware, leaving static scheduling a non-viable option.
Both dynamic  and guided scheduling have huge overhead associated 
(roughly 5 times that of static scheduling~\cite{openmpmeasure} )
making us look for more efficient options.
\begin{figure*}
\centering
\includegraphics[width=0.7\textwidth]{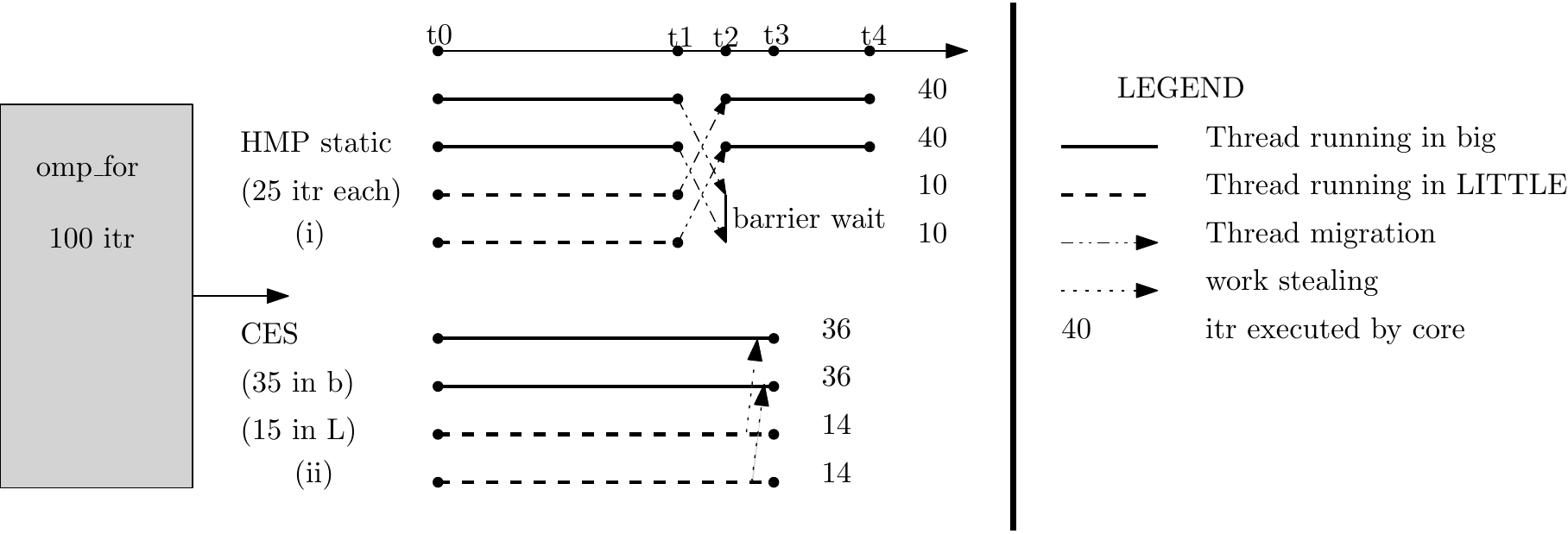}
\caption{ Figure comparing execution of \forconstruct{}. 
  The system demonstrated contains 2 \big{} and 2 \little{} cores. Numbers are symbolic.}
\label{fortheory}
\end{figure*}

Figure~\ref{fortheory} shows the inefficiency of static scheduling in HMP. 
Consider an \forconstruct{} with \niters=100 
executed with 4 parallel threads, on a system with 2 \big{} cores and 2 \little{} cores. 
The threads 
are allocated 25 iterations each. 
In HMP, once the threads that are initially scheduled in \big{} 
cores finish their work (time t1), they are migrated to \little{} while threads that are scheduled 
in \little{} are migrated to big (t2). 
Assume the \little{} cores have executed 10 iterations by time t1.  
The remaining 15 iterations in those threads will be executed by \big{} cores (t2 to t4). 
Meanwhile the \little{} cores do no meaningful work (the threads are waiting on barrier), 
leading to their underutilization, while big cores are overworked. 
With CES, we have a more balanced division of iterations based on the 
power of each core and the workload in each iteration. 
This increases effective parallelism, thereby reducing the execution time. 
To eliminate any imbalance in the initial division (due to inefficient modelling 
of runtime behaviour), CES also has a work-stealing framework built into it. 
A thread that has finished all its work can \emph{steal} 
work from other threads having  pending work. 
The \little{} cores consume lesser energy compared 
to big cores to execute the same code. 
In CES, \little{} cores take up relatively more work than in HMP,  thereby consuming lesser energy.


\begin{figure}
\centering
\begin{tabular}{ |@{~}l@{~}|@{~}l@{~}|}
\hline
A & \textbf{Finding victim thread to steal} \\
\hline
\begin{lstlisting}[style=rm]
1
2
3
4
5
6
7
8
9
\end{lstlisting}
&
\begin{lstlisting}[style=rm]
int getthread(){
  i = wl_with_max_itrs_left();
  if($\endworklist$[i]-$\iter$[i] > $\chunk$){
$\#pragma omp atomic write$
    $\status$[i] = $\statusshared$;  
    return i ;
  }
  return $\nthreads$;
}
\end{lstlisting} \\
\hline
\end{tabular}
\caption{Selection of victim thread for stealing}
\label{code:getthread}
\end{figure}

\begin{figure}
\centering
\begin{tabular}{ |@{~}l@{~}|@{~}l@{~}|}
\hline
B & \textbf{Worklist stealing-code} \\
\hline
\begin{lstlisting}[style=rm]
1
2
3
4
5
6
7
8
9
10
11
12
13
14
15
16
17
18
19
20
21
22
23
24
\end{lstlisting}
&
\begin{lstlisting}[style=rm]
int doitr(int t){
 if ( $\iter$[ t ] < $\endworklist$[ t ]){
  if($\status$[t] != $\statusshared$)
   return $\iter$[ t ] ++;
  else
   return $\lockupdate$($\iter$[t],$\iter$[t]+1,$\lock$[t]);
 }else {
$\#pragma omp atomic write$
   $\status$[t] = $\statusprivate$;
  int f = 1;
  do{
   int i = getthread ();
   if(i == $\nthreads$) return -1;
   int newend = $\endworklist{}$[i] - $\chunk$;
   int end = $\endworklist{}$[i];
   oe= $\lockupdate$($\endworklist$[i],newend,$\lock$[i]);
   if (oe == end ){
    $\endworklist$[ t ] = end ;
    return $\iter$[ t ]++;
   }else f = 0;
  } while (!f);
 }
 return -1;
}
\end{lstlisting}\\
\hline 
\end{tabular}
\caption{Do iteration code with stealing.}
\label{code:loopcommon}
\end{figure}

\begin{figure}
	\small
\centering
\begin{tabular}{ |@{~}l@{~}|@{~}l@{~}|@{~}l@{~}|@{~}l@{~}|  }
\hline
C & \textbf{ Initial Loop}  & D & \textbf{After}\\
\hline
\begin{lstlisting}[style=rm]
1
2
3
4
5
6
7
8
9
\end{lstlisting}
&
\begin{lstlisting}[style=rm]
$\cparallel$
{
  ...
 $\cfor$
 $\forh${
   S(i);
 }
  ...
}
\end{lstlisting} 
&
\begin{lstlisting}[style=rm]
1
2
3
4
5
6
7
8
9
10
11
12
13
\end{lstlisting}
&
\begin{lstlisting}[style=rm]
$\cparallel$
 {
  ...    
  initialize();
$\#pragma omp barrier$
  do{
   i = doitr(tid);
   S(i);
  }while(i != -1)
$\#pragma omp barrier$
  update_$\scaledendi$();
   ...
}
\end{lstlisting} \\
\hline
\end{tabular}
\caption{Transformations for \forconstruct.}
\label{loop}
\end{figure}

  

CES transforms \forconstruct{} to use a worklist based approach.
Each thread has a private worklist allocated with 
unequal chunk of iterations,
such that the  $\wlim$ is
minimal.
Once a thread finishes all iterations in its workload,
it first identifies the thread with
maximum iterations left (getthread function shown in Figure~\ref{code:getthread}) 
as the victim thread for stealing.
When a victim thread is selected for stealing, 
its worklist ceases being private and all operations
done on the worklist should be protected, by taking 
a \lock{} on the worklist.
\lockupdate\texttt{(var,val,lock)}  (B.6 and B.16 in Figure~\ref{code:loopcommon}) 
acquires \texttt{lock} 
and proceed to write \texttt{val} to \texttt{var} and returns 
the old value in \texttt{var}.
The data structure \status{} is maintained,
to denote current state of the worklist.
The value 
\statusshared{} (set when worklist  is selected for stealing, line
A.5 in Figure~\ref{code:getthread}) 
denotes shared state of the worklist and in other 
two stages (\statusprivate{} and \statusinitialprivate{}) the worklist is 
considered to be private.

The proposed transformation is shown in Figure~\ref{loop}.
The \texttt{initialize} function initializes \endworklist{} and \iter{} 
based on \scaledendi{} and also other variables used for transformation.
The \doiter{} function (shown in Figure~\ref{code:loopcommon}), 
returns the next iterations to be performed by the thread.
If the worklist is empty it 
proceeds to stealing
(lines B.8-24). It returns -1, when there are no iterations
available for the thread, on which the thread will stop working and
start waiting at the barrier (line D.10) for other threads.  

The variables used in the transformation are 
(i) \endworklist{}: this array keeps the last iteration 
in the worklist of the thread.
ii) \iter{}: This array keeps track of the current iteration 
that is being executed by each thread.
iii) \scaledendi{}: This array is used to set the initial division
of iterations. It will contain a scaled end point and
will need to be multiplied by \niters{} for actual end points.
If the calculated ratio depends on a variable whose value 
is  known only at runtime, the \scaledendi{} will be
replaced by an equation which will be evaluated at runtime.
iv) \chunk{}: This value decides the chunk\_size of iterations
for stealing based on stealing cost and iteration cost.
During execution, if only one team of threads is live 
(no nested parallelism) at a given point of time,
 all data structures can be made global (except 
\scaledendi{} created for $i^{th}$ \forconstruct{}).

If the \forconstruct{} has a chance of being
re-executed, we update the initial division for iterations based on
the current execution (the number of iterations executed by each thread), 
such that we will have a more balanced
division of $\wl{}$ on re-entry (D.11).

If an iteration has very little work (around 2-3 instructions), 
the stealing
cost would dominate the amount of work stolen thereby 
reducing the advantage we get from stealing.
We could  increase the value of \chunk{}
stolen by the stealer thread at a time, but increasing
the stealing chunk-size beyond a point would remove the 
balancing effect of the stealing process.
We discard the stealing logic for such loops
and proceed the execution with the initial division
of worklist. 
We categorize these loops as \emph{fixed size loop}.
This static division could lead to slight increase
in execution time and/or energy consumption, 
if the division is not balanced.

\subsection{Scheduling \secconst{}}

In \secconst{}, the default scheduling is arbitrary, which 
could result in inefficient scheduling and unnecessary migrations on \biglittle{}. 
A big \singlesection{} might be scheduled in \little{} core while \big{} cores
are scheduled with smaller workload. As a result the HMP scheduler will later 
introduce  thread migrations.

In CES the sheduling of \secconst{} is divided into two stages.
In the \emph{affinity-allocation}  stage, suitability
of each core for a \singlesection{} is determined using 
the type of operations performed in it (termed as affinity).
Based on the affinity, the \singlesection{} is allocated to the
core in \big{} or \little{} set with least workload.

Once the affinity allocation is complete, CES proceeds 
to \emph{normalization stage}, where
we try to balance the $\wl{}$ in each core. 
From the core with highest estimated $\wl{}$,
a \singlesection{} which shows least affinity towards 
it is
selected as the target-section.
The scheduler then allots the target-section 
to the core with least $\wl{}$.
If the new allocation configuration has a lower
$\wlim$, we select the new allocation, else 
we go ahead with the current allocation.

To implement this preferred scheduling the \secconst{} is 
transformed into an \forconstruct{}
with static scheduling, with the \niters{} set to
maximum \singlesection{} allocated to a thread multiplied by \nthreads{}. 
Each iteration would be given a \singlesection{} or nothing.
If the  \parseg{} is \singleconstruct{}, then it is taken as a special case of \secconst{}
with just one section for analysis.

\subsection{Thread Migration}
For the rest of the \parseg{} the amount of code executed by individual threads
in team will not change due to scheduling.
To reduce the $\wlim$ in these regions we introduce
\emph{thread-exchange} between \big{} and \little{} cores to reduce the imbalance.

We identify \emph{migration-points} (denoted as \texttt{mp} 
in line E.4, Figure~\ref{code:switching}) in code, as points with
the lowest thread exchange cost ($c_{ex}$). 
The $c_{ex}$ is calculated  
using the amount of live variables at that point, and \biglittle{} thread migration cost. 
On reaching the migration-point, the thread currently allocated in \little{} 
core ($attacker$ thread) will identify a thread in big core which has passed the
\emph{minimum-guarantee} point (denoted as \texttt{mgp} in line E.6) in its execution 
as the \emph{victim-thread}. 
The minimum-guarantee point is selected such that the $\wll$
for remaining code of the victim thread
is not larger than $\wlb$ for the remaining code of the attacker. 
Now if the estimated reduction in $\wlim$ is larger than the $c_{ex}$ the thread-exchange
is made.
When the ratio of  \little{} and \big{} cores available is 
more than one we might require more than one migration-point.
Each migration-point will have a paired minimum-guarantee point.
In the figure \texttt{pt}(line F.6) represents the migration point id, and the 
victim-thread should have reached the corresponding minimum guarantee-point
for the migration to happen (line F.7).

\begin{figure}
\centering
\begin{tabular}{ |@{~}l@{~}|@{~}l@{~}|@{~}l@{~}|@{~}l@{~}| }
\hline
\multicolumn{4}{|c|}{ \textbf{Thread switching transformation}} \\
\hline 
E & \textbf{Before} & F & \textbf{After} \\
\hline
\begin{lstlisting}[style=rm]
1
2
3
4
5
6
7
8










\end{lstlisting}
&
\begin{lstlisting}[style=rm]
$\textbf{\#pragma omp parallel}$
{
  Si;
  //mp
  Sj;
  // mgp
  Sk;
}

















\end{lstlisting}
&
\begin{lstlisting}[style=rm]
1
2
3
4
5
6
7
8
9
10
11
12
13
14
15
16
17
18
19
\end{lstlisting}
&
\begin{lstlisting}[style=rm]
$\textbf{\#pragma omp parallel}$
{
 Si; i=0;
if(inLITTLE()){
 while(i<$\texttt{\nthreads}$){
  if(mg[i]>=pt
   &&inbig(i))
   migrate(tid,i);
   break;
 }
 i++;
}
 Sj;
 if(inbig(tid)){
   $\textbf{\#pragma omp atomic}$
   mg[tid] = i;
  }
  Sn;
}
\end{lstlisting}\\
\hline
\end{tabular}
\caption{Transformation for thread switching. in\little() and inbig()
is to check core running current thread.}
\label{code:switching}
\end{figure}

\section{Implementation}

We use IMOP~\cite{imop}, a source to source \omp{} compiler. 
The transformed code will also be in OpenMP and
can be compiled using compilers like gcc. 
The compiler framework is divided into two phases, the analysis phase and transformation
phase. 
Pre-processing includes removal of macros, function in-lining whenever possible
and also MHP analysis for division of \parseg{}. 
The analysis phase involves generating program metrics, and 
balancing the $\wl{}$ on each thread.
The transformation phase implements the transformations
based on the suggestions made during the analysis phase.
  

\begin{table*}[ht]
\begin{center}
\caption{Excution time and Energy consumption for different configurations of big and little.
Time in seconds and Energy in Joules. 4 b 4 L stands for 4 \big{} and 4 \little{} cores.
The \little{} are ran at 1.3 GHz while \big{} are ran at 1.9GHz.}
\label{results}
\begin{tabular}{| c| c| c| c| c | c| c| c| c | c| c| c| c|}
\hline
 BM &   \multicolumn{4}{|c|}{4 b 4 L} & \multicolumn{4}{|c|}{2 b 4 L } & \multicolumn{4}{|c|}{2 b 2 L} \\ 
\hline
 & \multicolumn{2}{|c|}{hmp} & \multicolumn{2}{|c|}{ces} & \multicolumn{2}{|c|}{hmp} 
 & \multicolumn{2}{|c|}{ces} & \multicolumn{2}{|c|}{hmp} & \multicolumn{2}{|c|}{ces} \\
\hline
 & Time & Energy& Time & Energy& Time & Energy& Time & Energy& Time & Energy& Time & Energy   \\ \thickline
\texttt{EP A} & 15.05 & 58.06  & 13.07 & 55.09  & 22.06  & 51.68   & 19.09  & 49.09  & 29.05  & 61.64   & 19.66 & 56.06  \\
\texttt{EP B} & 58.17 & 221.31 & 51.19 & 207.03 & 86.02  & 203.65  & 75.23  & 186.19 & 114.80 & 244.14  & 77.08 & 225.40 \\
\texttt{CG}   & 5.07  & 43.18  & 3.16  & 29.11  & 6.06   & 36.61   & 4.09   & 24.24  & 5.57   & 32.61   & 4.8   & 26.77  \\
\texttt{IS}   & 1.62  & 6.53   & 1.60  & 6.44   & 1.44   & 4.52    & 1.42   & 4.51   & 1.4    & 4.02    & 1.38  & 4.09 \\
\texttt{sec}  & 58.52 & 272.68 & 59.78 & 151.9  & 96.7   & 236.85  & 59.8   & 165.65 & 117.05 & 291.37  & 76.25 & 253.50 \\

\hline
\end{tabular}
\end{center}
\end{table*}
\begin{figure*}[t]
\centering
\includegraphics[width=0.9\columnwidth]{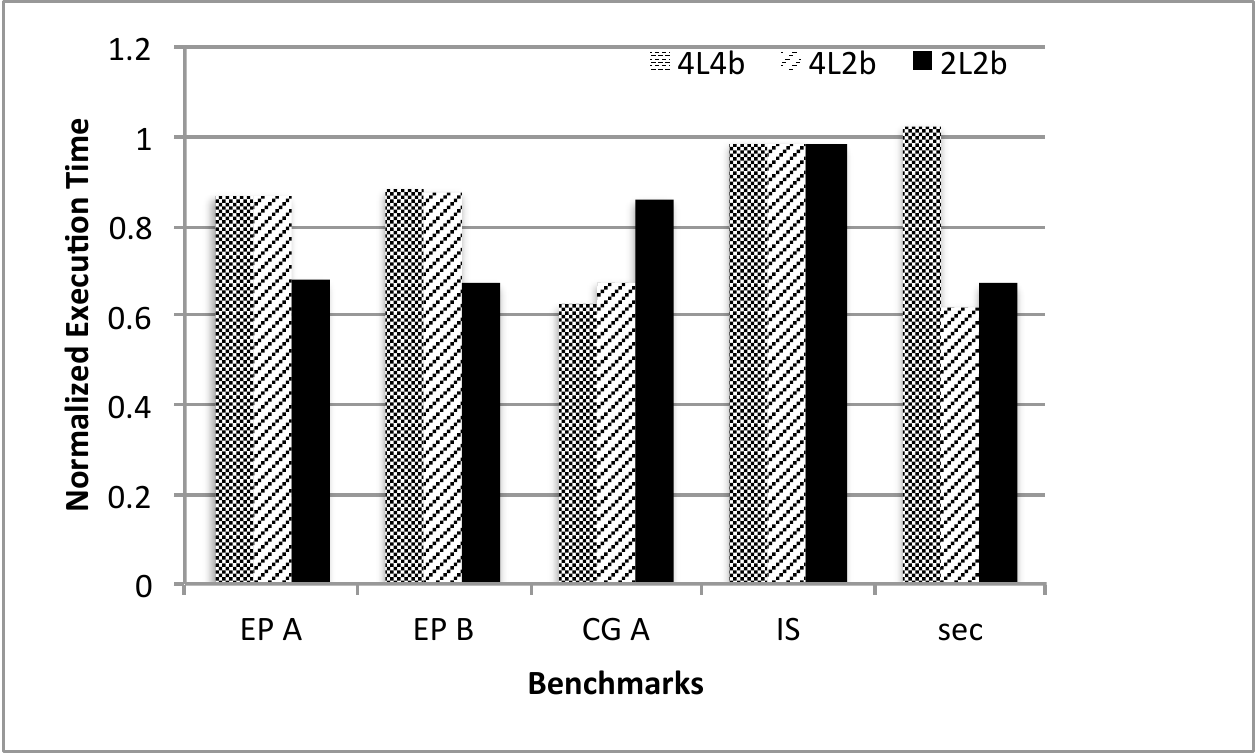}
\includegraphics[width=0.9\columnwidth]{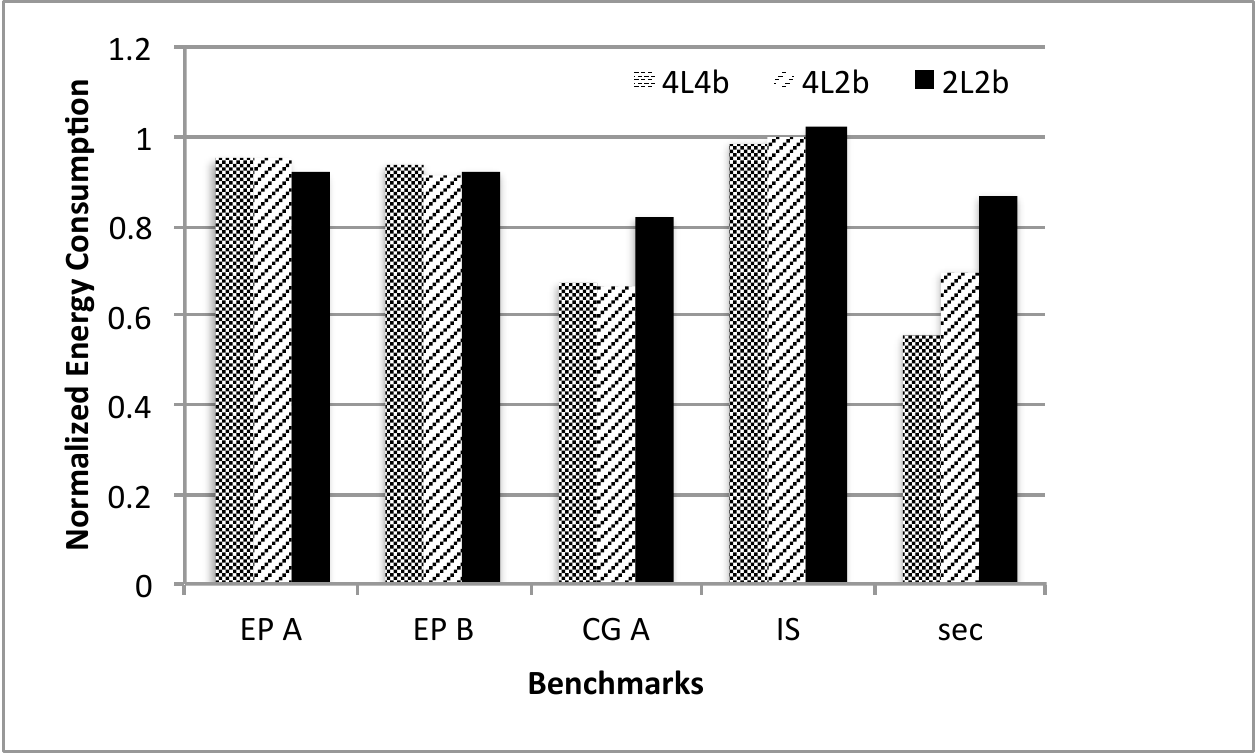}
\caption{ Normalized execution time and energy consumption running \little{} at 1.3 GHz
and \big{} at 1.9GHz.4L4b stands for 4 \little{} cores and 4 \big cores{}.}
\label{plots:lb}
\end{figure*}

\begin{figure*}
\centering
\includegraphics[width=0.9\columnwidth]{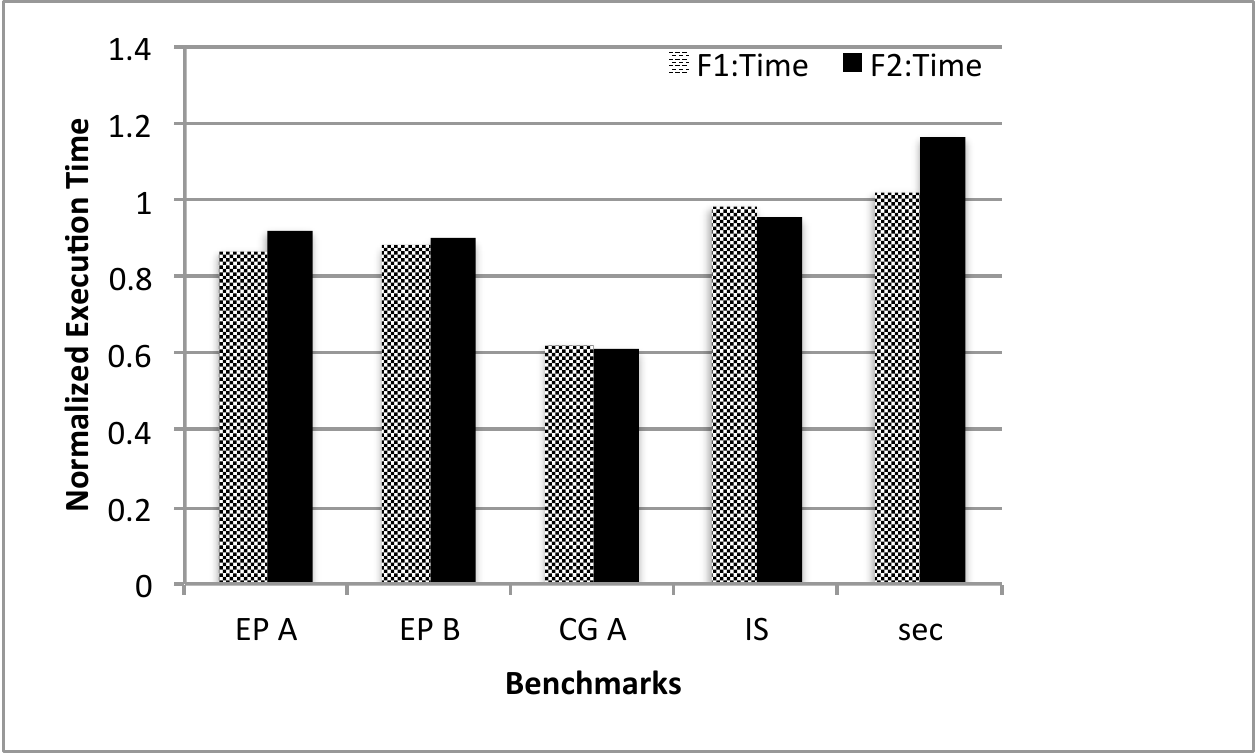}
\includegraphics[width=0.9\columnwidth]{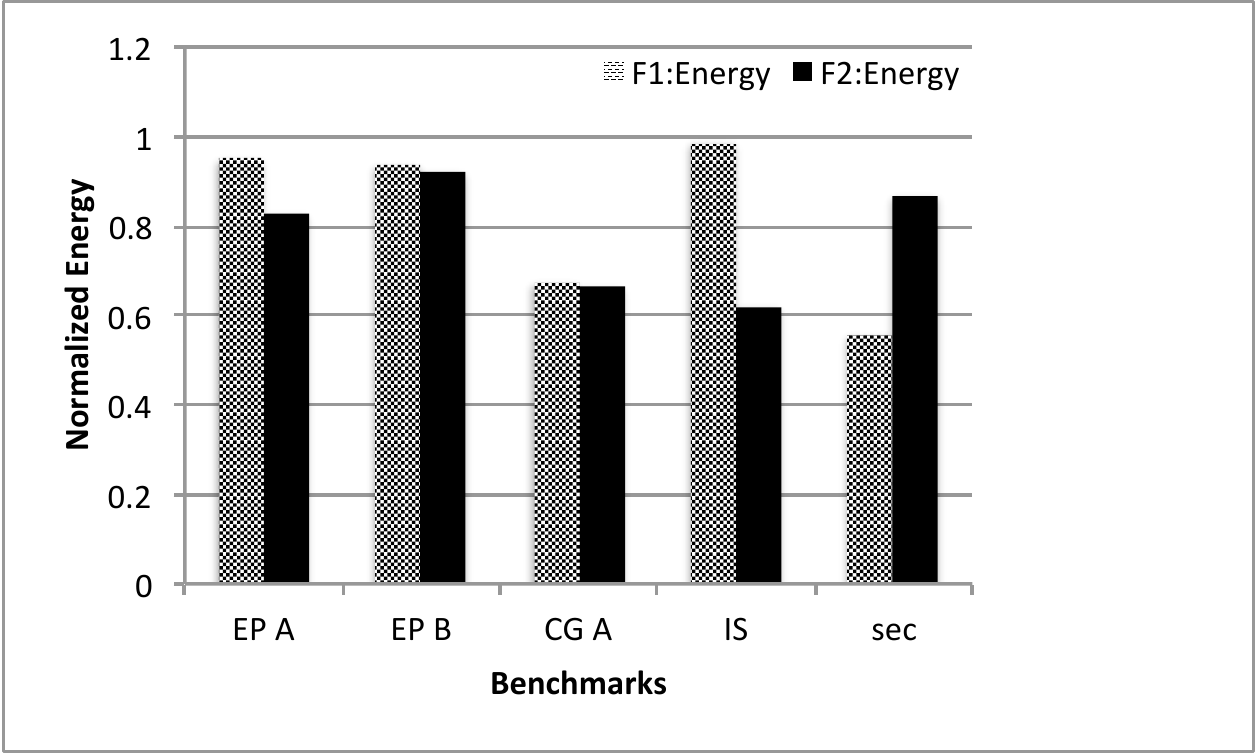}
\caption{ Normalized execution time and energy consumption for different configurations. For f1
we ran \little{} at 1.3GHz and for f2 \little{} are ran at 1GHZ. \big{} were ran
at 1.9GHz for both. System with 4 \little{} and 4\big{} cores.}
\label{plots:f1f2}
\end{figure*}


\section{Results}

Results of our transformations on standard NPB benchmarks~\cite{npb} are explained here.
We have set up a Linux environment (Ubuntu 14.04) enabled with HMP scheduling in 
Odroid-XU3 board~\cite{odroid} installed with Samsung Exynos 5422~\cite{exynos}. 
Exynos 5422 implements \emph{big.LITTLE} technology with 4 little and 4 big cores.

Table~\ref{results} shows execution time and overall system energy consumption of each benchmark. 
Readings shown here are the average of 5 runs with \little{} cores  running at 1.3Ghz and \big{} 
cores running
at 1.9GHz.
We have tried our compiler in three different hardware
combinations : (i) 2 \little{} and 2 \big{} cores (ii) 4 \little{} and 2 \big{} core (iii) 4 \little{} and 4 \big{} cores.
Figure~\ref{plots:lb} shows normalized execution time and energy consumption for
the different hardware combinations.
It shows on an average 18\% reduction in execution time
and 14\% reduction in energy consumption.

\texttt{EP} has a single \forconstruct{} which has been transformed using a worklist with stealing. 
\texttt{EP} has been studied in two sizes (\texttt{A} \& \texttt{B}) to find the effect of work-stealing on performance. 
For both sizes we get on an average, 20\% improvement in runtime and 8\% improvement in 
energy consumption showing the 
stealing algorithm scales well with size.   
Benchmarks \texttt{CG} and \texttt{IS} are made of small \forconstruct{} (one to two instructions).
These are transformed to simple static worklists.
Changing ratio of \big{} and \little{} cores available have low effect on performance improvement
with the CES able to adapt to the changes in hardware configurations.

We have also taken a modified version of multitask benchmark from FSU-\omp{} benchmark (denoted
by \texttt{sec} in the tables). 
The original program had two \singlesection{} (one computing
prime and another sine) which was copied 4 times
(since maximum \nthreads{} = 8) to make the modified benchmark.
Prime \singlesection{} showed more affinity towards big.
In HMP, arbitrary scheduling of sections caused variations 
in execution time based on 
the initial scheduling, which was more evident when \nthreads{} was less than the number of 
\singlesection{}.
By obliging to section \emph{affinity} of \singlesection{} 
we are able to gain huge energy benefits with little
effect on the execution.

We have also tested effectiveness of CES, for different frequencies for \biglittle.    
Figure~\ref{plots:f1f2} shows normalized execution time and energy consumption for two frequency
configurations. 
Readings of $f1$ are obtained while running \big{} at 1.9GHz
and \little{} at 1.3GHz while that of  $f2$ are obtained while running \big{} at 1.9GHz
and \little{} at 1GHz.
For $f1$ on an average, we get 16\% improvement in execution time
and 23\% reduction in energy consumption. For $f2$ we get 11\% reduction in runtime
and 24\% reduction in energy consumption.  
Energy gain for \texttt{IS} in $f2$  is higher (28\% compared 
to 3\% in $f1$) since the workload division
obtained was more balanced when compared to what we got with $f1$.

\section{Related Work}


Most of the scheduling techniques proposed are run-time in nature
~\cite{pie_VanCraeynest,tongli,Chronakicriticality,Becchi}. 
In PIE (Performance Impact Estimation)~\cite{pie_VanCraeynest},
hardware counters for CPI, ILP and MLP are used to estimate the performance that can be  achieved,
if the thread was to run in other type of core at run-time. 
The scheduling is then decided based on the estimate. 
Asymmetric Multi Processor 
Scheduling (AMPS) 
~\cite{tongli} is another
run time scheduler which quantifies the computing power of 
each core based on the weakest core (scaled power). The scheduling 
policy is structured to balance the workload based on each core's scaled power. 
 Profiling is used to 
find out the affinity of each thread efficiently to schedule the threads to core which shows~\cite{biasKoufaty} 
maximum affinity and improve the efficiency of the system.
The works above do not consider the nature and type of multithreaded programs that are run,
and the run-time methods proposed are based on the current state of the system. 

Scheduling algorithms described in ~\cite{DuBoisCriticality,Chronakicriticality, criticalaater} 
that find threads and program sections that are critical in nature, and schedule those 
in big cores.  In ~\cite{Chronakicriticality} the critical
sections are identified at run-time, by prioritizing those tasks that form 
part of the longest path in dynamic task dependency graph.  
There are also scheduling algorithms which are aimed at real time 
applications in a QoS (Quality of Service) - performance 
trade off~\cite{approximationchen}.
These works show that knowledge 
about nature of workload can help in optimizing the performance.
Compiler analysis and transformations are heavily relied upon, in these
works, for optimization.

\section{Conclusion}
This paper explain a hardware aware compiler for \omp{} in \biglittle{}.
The compiler directed scheduling can
help balancing
the future workload in 
each thread to reduce the runtime and power consumption in runtime. 
The present day hardware-unaware compiler assumes SMP and the scheduling
leads to unfair division of workload.
Our method shows great promise with around 18\% improvement in execution time 
on an average  14\% improvement in energy consumption.
Our future focus involves providing a runtime framework  to handle hardware
heterogeneity, finding the best hardware configuration for \biglittle{} 
in terms of power and runtime.

\bibliographystyle{plain}

\balancecolumns
\end{document}